# Probing the spin polarization of current by soft X-ray imaging of current-induced magnetic vortex dynamics

Shinya Kasai[1], Peter Fischer[2], Mi-Young Im[2], Keisuke Yamada[1], Yoshinobu Nakatani[3], Kensuke Kobayashi[1], Hiroshi Kohno[4] & Teruo Ono[1]

[1]*Institute for Chemical Research, Kyoto University, Uji 611-0011, Japan*

[2] *Center for X-Ray Optics, Lawrence Berkeley National Lab, 1 Cyclotron Road, Berkeley, California 94720, USA*

[3]*University of Electro-communications, Chofu 182-8585, Japan*

[4]*Graduate School of Engineering Science, Osaka University, Toyonaka 560-8531, Japan*

## Abstract

Time-resolved soft X-ray transmission microscopy is applied to image the current-induced resonant dynamics of the magnetic vortex core realized in a micronsized Permalloy disk. The high spatial resolution better than 25 nm enables us to observe the resonant motion of the vortex core. The result also provides the spin polarization of the current to be $0.67 \pm 0.16$ for Permalloy by fitting the experimental results with an analytical model in the framework of the spin-transfer torque.

The manipulation of magnetisation by currents is a key technology for future spintronics because it does not require the application of an external magnetic field to control the devices [1]. The current-driven magnetic domain wall (DW) motion [2-4] in ferromagnetic nanowires is currently considered to be one of the promising candidates for realizing new spintronic devices such as magnetic racetrack memory [5]. The physics underlying the current-driven DW motion is the so-called 'spin-transfer torque', first proposed by Berger [6]. When a spin-polarized current passes through a DW, it exerts a torque on the magnetisation through a microscopic $s$-$d$ exchange interaction. This torque rotates the magnetisation vector in the DW and drives the DW in the direction of the electron flow. The strength of the spin-transfer torque is proportional to the spin-current density, $j_s$, and hence, to the degree of the spin-polarization of the current, defined by $P = j_s/j$, where $j$ is the current density. Therefore, a precise knowledge of the value of $P$ is essential for controlling ferromagnetic nanostructures by electric currents. So far, this value has been estimated from transport measurements such as tunnelling spectroscopy [7], Andreev reflection [8] and giant magnetoresistance measurements [9]. However, all these methods involve the interface of magnetic and nonmagnetic materials, whose effects are not easy to discriminate. For example, in permalloy, the value of $P$ estimated by these methods ranges from 0.4 to 0.7.

In this paper, we show for the first time that by imaging the current-driven resonant vortex dynamics using high-resolution soft X-ray microscopy, one can experimentally determine the value of $P$ by the direct use of the spin-transfer torque. This method is not affected by interfaces, and is expected to directly probe the bulk and intrinsic properties of the ferromagnetic material.

The magnetic vortex core confined in a ferromagnetic circular disk [10, 11] exhibits a characteristic excitation called translational mode, which [12-15] corresponds to the circular motion around the disk centre. This circular motion can be resonantly excited by an ac current [16-18], and its radius is proportional to the spin-current density, and thus, to $P$. Thus, the measurement of this radius offers a direct method to determine $P$.

We have performed time-resolved magnetic transmission soft X-ray microscopy (M-TXM) [19, 20] to obtain the real-time and real-space imaging of the current-induced translational motion of the vortex core with a spatial resolution that is better than 25 nm and a time resolution of 70 ps. The high spatial resolution enables us for the first time to determine both the spin polarization of the current and Gilbert damping parameter through an analytical model calculation coupled with micromagnetic simulations.

Figure 1 shows a scanning electron microscope image of the sample and the schematics of the experimental setup for time-resolved M-TXM at beamline 6.1.2 at the Advanced Light Source (ALS) in Berkeley, CA. The samples were prepared on 200-nm-thick $Si_3N_4$ membrane substrates to allow for sufficient transmission of the soft X-rays. Permalloy ($Fe_{19}Ni_{81}$) disk structures with diameters of 1.5 μm and thicknesses of 40 nm were defined by a combination of electron beam lithography with standard lift-off techniques. Au electrodes (thickness: 80 nm) form a coplanar waveguide structure for applying the ac excitation current.

The optical setup of the X-ray microscope and X-ray magnetic circular dichroism effect used as magnetic contrast are described elsewhere [19, 20]. Fresnel zone plates provide high spatial resolution, and a stroboscopic pump-probe scheme is utilized for time-resolved studies, where a time resolution of 70 ps is set by the width of the X-ray pulses emitted from a bending magnet in the synchrotron. The X-ray images were recorded at the Ni $L_3$ absorption edge at a photon energy of 854 eV with the ALS operating in a 2-bunch mode. In this mode, the X-ray pulses have a repetition frequency of 3 MHz, i.e. a 328-ns time distance between subsequent probe pulses; therefore, the phase of the pumping pulse, i.e. the ac excitation currents (or motion of the vortex core) has to be synchronized with this clock signal. Using an arbitrary function generator (AFG), a pulse-modulated ac voltage of 100 ns in duration was applied to the circular disk to excite the motion of the magnetic vortex core. The reflection signal at the sample was monitored through a directional bridge by using a real-time oscilloscope to evaluate the excitation current density and temperature increase during the measurement. As the temperature of the sample was estimated to be approximately 380 K under the present experimental conditions, the effect of Joule heating can be neglected in our experimental data. The excitation frequencies were varied from 150 MHz to 250 MHz to cover the resonant frequency of the translational motion of the magnetic vortex core of the sample.

Figure 2(a) shows a sequence of the time-resolved M-TXM images of the sample. The excitation current density ($j$) is $1.3 \times 10^{11}$ A/m$^2$, and the excitation frequency ($f$) is 220 MHz, which is close to the resonance frequency of 218 MHz derived from micromagnetic simulations. In Fig. 2, $t$ is the probing time, with $t = 0$ set to 80 ns after starting the pumping operation by the ac excitation current. At $t = 0$, the vortex core is located slightly below the centre of the disk. The position of the vortex core is shifted leftwards at $t = 1$ ns. Then, the vortex core follows a clockwise circular trajectory and returns to its initial position at $t = 5$ ns. These results confirm a circular motion of the vortex core with a period of approximately 5 ns, which is consistent with our excitation frequency (220 MHz) within an experimental accuracy. The clockwise rotation of the vortex core can be clearly observed, which is consistent with the downward core magnetisation, $m_z = -1$, or polarity $p = -1$, determined by magnetic force microscopy

An analytical treatment of the vortex core motion is possible by using the Thiele equation

$$\vec{G} \times \vec{v} + \alpha D \vec{v} + k \vec{X} - \vec{G} \times \vec{u} = 0, \quad (1)$$

extended to include the spin-transfer torque [21,22]. The first term represents the gyrotropic force (gyroforce) derived from the time derivative of the magnetisation. The gyrovector $\vec{G}$ is normal to the plane of the disk; hence, the gyroforce acts as a transverse force to the core velocity $\vec{v}$. The second term represents the damping, with $D$ being a diagonal component of the dissipation dyadic determined by the magnetisation distribution. The third term is a restoring force when the position $\vec{X}$ of the vortex core is displaced from the equilibrium position (centre of the disk). The force constant $k$ is defined by $k\vec{X} = -\partial W / \partial \vec{X}$, where $W$ is the total magnetic energy of the system. The last term is an equivalent force that arises due to the spin-transfer torque. The vector $\vec{u} \left( = -P\vec{j} / 2eM_S \right)$ is the drift velocity of the spin, where $e$ is the elementary charge and $P$ is the spin polarization of the current.

By solving eq.(1) with $\vec{u} = u_0 \cos(2\pi f t) \vec{e}_x$, where $\vec{e}_x$ is a unit vector in the direction of the current, the radius of the circular motion of the core is obtained as

$$a(f) = \frac{1}{4\pi} \frac{u_0}{\sqrt{(f - f_0)^2 + (\hat{\alpha} f)^2}}, \quad (2)$$

where $f_0 = k / 2\pi G$ is the resonance frequency of the translational mode of the core, and $\hat{\alpha} = \alpha D / G$ is the effective damping parameter (see Supplementary Discussion). The radius $a(f)$ is peaked at $f = f_0$, with a peak value of $u_0 / 4\pi \hat{\alpha} f_0$, and a half-width at half-maximum (HWHM) $w = \sqrt{3}\hat{\alpha}f_0$. Once the ratio $D/G$ is provided, $P$ and $\alpha$ can be experimentally estimated through these analytical expressions. For the present sample, $D/G$ was estimated to be 2.7 by micromagnetic calculations using typical material parameters for permalloy: $\mu_0 M_s = 1$ T for the saturation magnetisation and $A = 1.0 \times 10^{-11}$ J/m for the exchange stiffness constant. Now, we can estimate both $P$ and $\alpha$ by fitting our experimental data in eq. (2); Figure 3 summarizes the radius $a(f)$ estimated from the experimental results (open circles) and a fitting curve based on eq. (2). The parameters obtained by the fitting are $P = 0.67 \pm 0.16$, $\alpha = 0.01 \pm 0.003$ and $f_0 = 219.3$ MHz.

We have also carried out micromagnetic simulations in the framework of the Landau-Lifshitz-Gilbert equation including a spin-transfer term [16, 18]. Figure 2(b) shows a sequence of the magnetisation distribution obtained from these micromagnetic simulations with $f = 220$ MHz, $P = 0.67$, $j = 1.3 \times 10^{11}$ A/m² and $\alpha = 0.01$. The results clearly reproduce the resonant motion of the vortex core obtained in our experiment.

In summary, we have used M-TXM with high spatial resolution to record the resonant motion of the magnetic vortex core, and determined the spin polarization, $P$, of the current. In contrast to previous measurements, the present method does not involve interface effects [2–4], and is expected to probe the intrinsic property of the materials. The obtained value will be suitably used to analyze the current-driven motion of other systems such as DWs.

We thank the staff of CXRO and ALS for their assistance with the experiment at BL 6.1.2. This study was partly supported by the MEXT Grant-in-Aid for Scientific Research in Priority Areas and JSPS Grant-in-Aid for Scientific Research(S). The operation of the X-ray microscope was supported by the Director, Office of Science,

Office of Basic Energy Sciences, Materials Sciences and Engineering Division, U.S.

Department of Energy.

[1] C. Chappert, A. Fert, and F. N. V. Dau, *Nature Mater.* **6**, 813 (2007).

[2] A. Yamaguchi, T. Ono, S. Nasu, K. Miyake, K. Mibu, and T. Shinjo, *Phys. Rev. Lett.* **92**, 077205 (2004).

[3] M. Yamanouchi, D. Chiba, F. Matsukura and H. Ohno, *Nature* **428**, 539–542 (2004).

[4] M. Hayashi, *et al.*, *Phys. Rev. Lett.* **96**, 197207 (2006).

[5] S. S. P. Parkin, *US Patent* 309,6,834,005 (2004).

[6] L. Berger, *J. Appl. Phys.* **55**, 1954 (1984).

[7] D. J. Monsma, and S. S. P. Parkin, *Appl. Phys. Lett.* **77**, 720 (2000).

[8] R. J. Soulen *et al.*, *Science* **282**, 85 (1998).

[9] J. Bass and W. P. Pratt Jr, *J. Magn. Magn. Mater.* **200**, 274 (1999).

[10] T. Shinjo, T. Okuno, R. Hassdorf, K. Shigeto, and T. Ono, *Science* **289**, 930 (2000).

[11] A. Wachowiak *et al.*, *Science* **298**, 577 (2002).

[12] K. Yu. Guslienko *et al.*, *J. Appl. Phys.* **91**, 8037 (2002).

[13] J. P. Park, P. Eames, D. M. Engebretson, J. Berezovsky and P. A. Crowell, *Phys. Rev. B* **67**, 020403(R) (2003).

[14] V. Novosad *et al.*, *Phys. Rev. B* **72**, 024455 (2005).

[15] B. Van Waeyenberge *et al.*, *Nature* **444**, 461 (2006).

[16] S. Kasai, Y. Nakatani, K. Kobayashi, H. Kohno, and T. Ono, *Phys. Rev. Lett.* **97**, 107204 (2006).

[17] S. Kasai, Y. Nakatani, K. Kobayashi, H. Kohno and T. Ono, *J. Magn. Magn. Mater.* **310**, 2351 (2007).

[18] K. Yamada *et al.*, *Nature Mater.* **6**, 270–273 (2007).

[19] P. Fischer, *Current Opinion in Solid State and Materials Science* **7**, 173-179 (2003).

[20] P. Fischer *et al.*, *J. Magn. Magn. Mater.* **310**, 2689–2692 (2007).

[21] A. Thiaville, Y. Nakatani, J. Miltat and Y. Suzuki, *Europhys. Lett*. **69**, 990–996 (2005).

[22] J. Shibata, Y. Nakatani, G. Tatara, H. Kohno and Y. Otani, *Phys. Rev. B* **73**, 020403(R) (2006).

Figure captions

Figure 1

Scanning electron microscope image of the sample and schematic illustration of the experimental setup for imaging the spin dynamics by time- and space-resolved magnetic soft X-ray microscopy of the vortex core motion in a permalloy disk (diameter: 1.5 μm; thickness: 40 nm). The current density and sample resistance upon applying an ac excitation current are estimated from the reflection signal monitored using the SWR bridge.

Figure 2

(a) Time-resolved sequence of MTXM images of the Permalloy disk sample. Images obtained at various delay times from $t = 0$ ns up to $t = 9$ ns show the translational motion of the core centre. Black (white) contrast indicates that the direction of the magnetization is to the left (right). (b) Numerical simulation results of the current-induced vortex translational motion for $j = 1.3 \times 10^{11}$ A/m$^2$, $f_{exc}$=220 MHz, $P = 0.67$ and $\alpha = 0.01$.

Figure 3

Radius, $a$, of the circular motion of the core as a function of the excitation frequency, $f$. The open circles represent the experimental results obtained with $j = 1.3 \times 10^{11}$ A/m$^2$, whereas the solid curve represents the fitting by eq. (2).

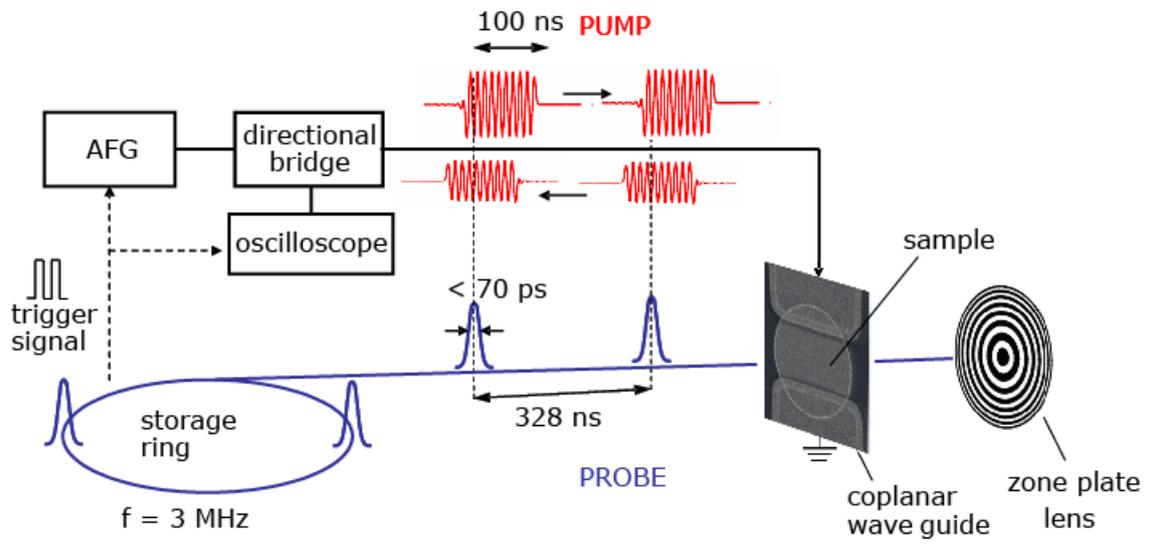

Figure 1

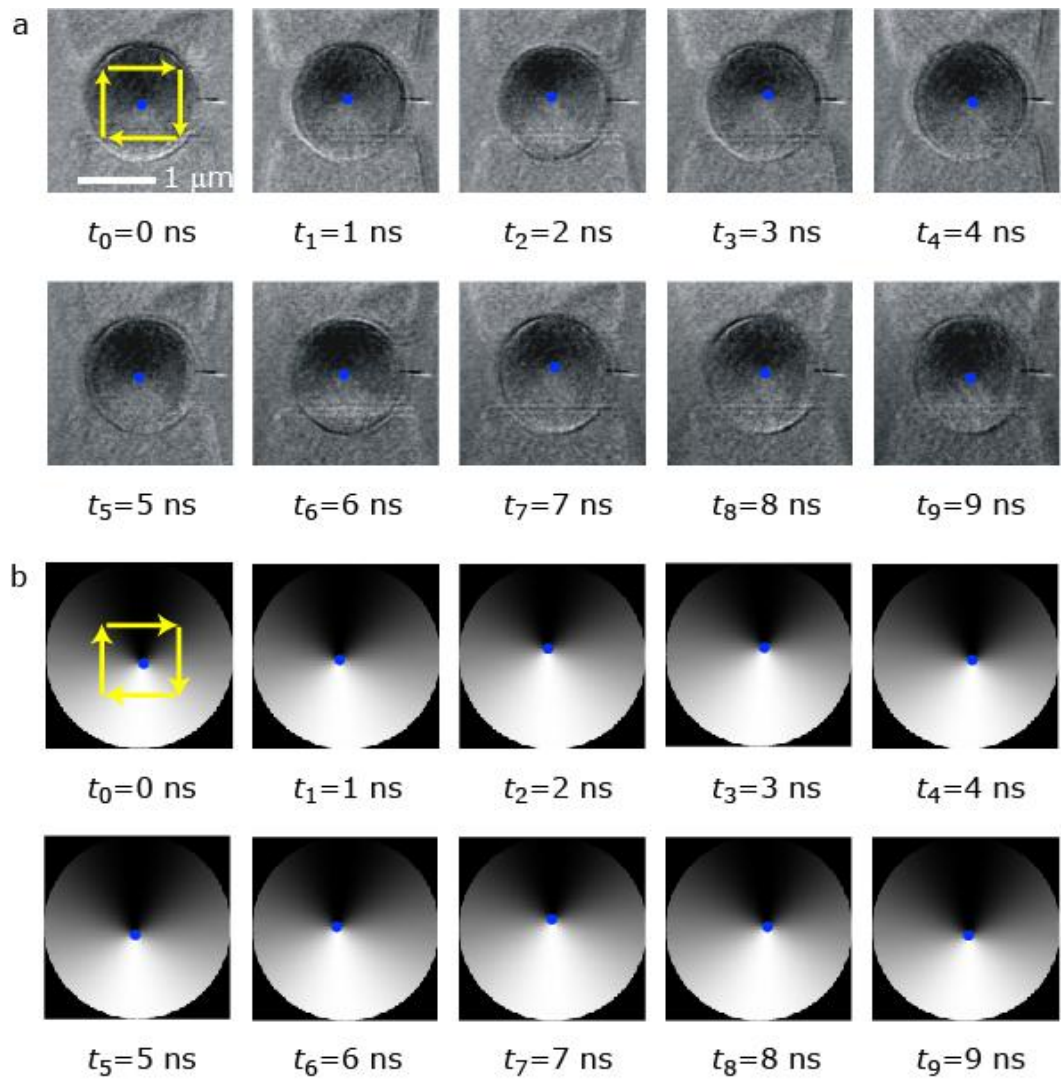

a

$t_0=0$ ns    $t_1=1$ ns    $t_2=2$ ns    $t_3=3$ ns    $t_4=4$ ns

$t_5=5$ ns    $t_6=6$ ns    $t_7=7$ ns    $t_8=8$ ns    $t_9=9$ ns

b

$t_0=0$ ns    $t_1=1$ ns    $t_2=2$ ns    $t_3=3$ ns    $t_4=4$ ns

$t_5=5$ ns    $t_6=6$ ns    $t_7=7$ ns    $t_8=8$ ns    $t_9=9$ ns

Figure 2

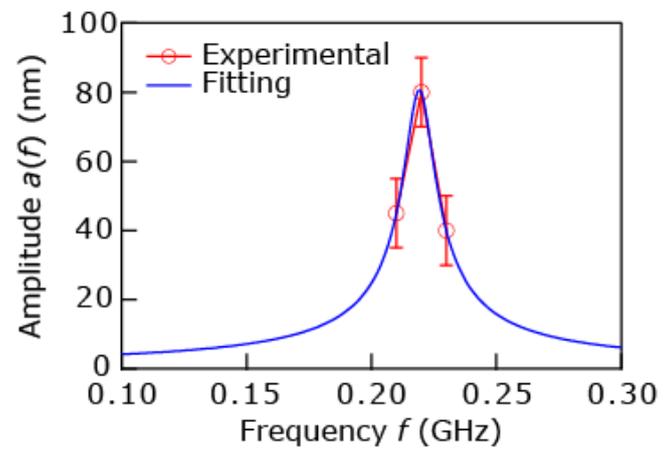

Figure 3